\documentclass[conference]{IEEEtran}
\IEEEoverridecommandlockouts
\usepackage{cite}
\usepackage{amsmath,amssymb,amsfonts}
\usepackage{algorithmic}
\usepackage{subcaption}
\usepackage{graphicx}
\usepackage{textcomp}
\usepackage{xcolor}

\usepackage{hyperref}
\usepackage{physics}
\def\BibTeX{{\rm B\kern-.05em{\sc i\kern-.025em b}\kern-.08em
    T\kern-.1667em\lower.7ex\hbox{E}\kern-.125emX}}
\hypersetup{
    colorlinks=true, 
    linkcolor=blue,  
    citecolor=blue, 
    urlcolor=red     
}

\title{Demonstration of Scalability and Accuracy of Variational Quantum Linear Solver for Computational Fluid Dynamics}

\author{
\IEEEauthorblockN{Ferdin Sagai Don Bosco\IEEEauthorrefmark{1}, Dhamotharan S\IEEEauthorrefmark{2}, Rut Lineswala\IEEEauthorrefmark{3}, Abhishek Chopra\IEEEauthorrefmark{4}}
\\
\IEEEauthorblockA{ BosonQ Psi (BQP) Corporation \\
New York, USA\\
\IEEEauthorrefmark{1} Senior Computational Scientist, Email: ferdindon@bosonqpsi.com\\
\IEEEauthorrefmark{2} Quantum Computing Developer\\
\IEEEauthorrefmark{3} Chief Technology Officer\\
\IEEEauthorrefmark{4} CEO and Chief Scientific Officer, Email: abhishek.chopra@bosonqpsi.com}
}

\begin{document}
\maketitle

\begin{abstract}
The solution for non-linear, complex partial differential Equations (PDEs) is achieved through numerical approximations, which yield a linear system of equations. This approach is prevalent in Computational Fluid Dynamics (CFD), but it restricts the mesh size since the solution of the linear system becomes computationally intractable when the mesh resolution increases. The reliance on the ability of High-Performance Computers (HPC) to scale up and meet these requirements is myopic; such very high-fidelity simulations require a paradigm shift in computing. This paper presents an exploration of quantum methodologies aimed at achieving high accuracy in solving such a large system of equations.

Leveraging recent works in Quantum Linear Solver Algorithms (QLSA) and variational algorithms suitable for Quantum Simulation in HPC, we aspire to push the boundaries of CFD-relevant problems that can be solved on hybrid quantum-classical framework. To this end, we consider the 2D, transient, incompressible, viscous, non-linear coupled Burgers equation as a test problem and investigate the accuracy of our approach by comparing results with a classical linear system of equation solvers, such as the Generalized Minimal RESidual method (GMRES). Through rigorous testing, our findings demonstrate that our quantum methods yield results comparable in accuracy to traditional approaches. Additionally, we demonstrate the accuracy, scalability, and consistency of our quantum method. Lastly, we present an insightful estimation of the resources our quantum algorithm needs to solve systems with nearly 2 billion mesh points.
\end{abstract}


\section{Introduction}
Partial Differential Equations (PDE) are a powerful mathematical tool for describing the complex phenomena observed in all aspects of the universe. Its relevance in Science is evidenced through the Maxwell equations \cite{taflove1975}, Schrodinger equation \cite{fibich2015}, Engineering, evident from the Fourier heat transfer \cite{narasimhan1999} and Navier-Stokes equations\cite{temam2001}, and Finance, embodied by the Black-Scholes equation\cite{barles1998} is undeniable. Active research is constantly underway to discover new PDEs, even with the aid of data-driven techniques, such as Physics-Informed Neural Networks (PINNs)\cite{xu2021}. 

However, describing a physical phenomenon is only the first challenge; solving it is even more daunting. This is often insurmountable, especially when the PDE features non-linearity (Burgers equation\cite{bonkile2018}), incomplete closure (Navier-Stokes equations), or is very complex, such as the integro-differential nature of the Boltzmann equation \cite{cercignani1988}.

Non-linear PDEs are specifically prominent in transport mechanics, with applications including plasma fields \cite{seadawy2017a}, nanofluid flow \cite{sheikholeslami2017}, and shallow water waves \cite{seadawy2017b}. Fluid dynamics and the associated field of Computational Fluid Dynamics (CFD) possess a rich ensemble of such PDEs, which capture complex physical phenomena, such as convection-diffusion, heat transfer, shock wave formation and propagation, etc. However, an analytical solution is only feasible for a grossly simplified version of the original phenomena, which is of little real-world importance. 

\begin{figure}[htbp]
\centering 
\includegraphics[width=\columnwidth]{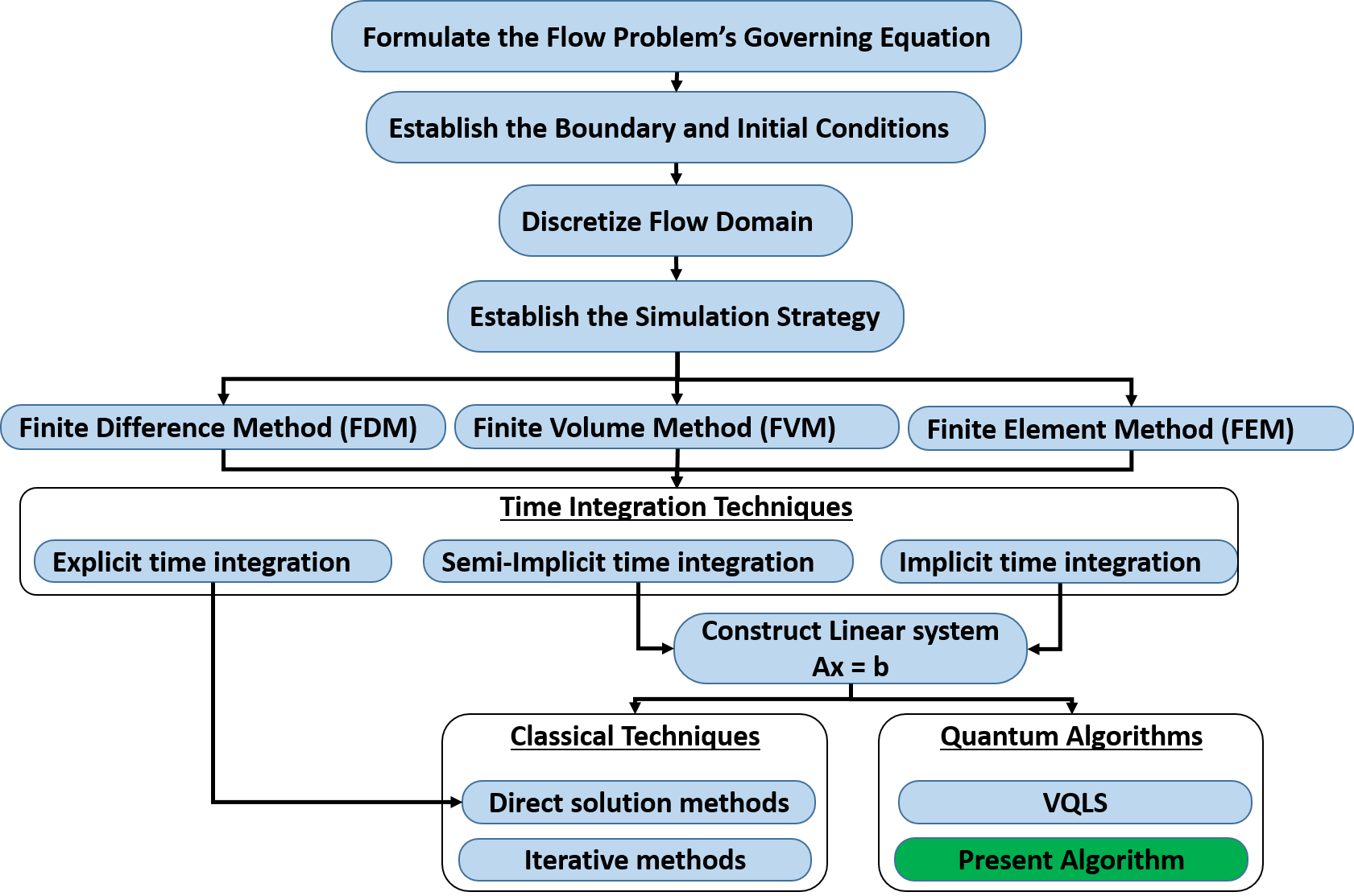}\quad
\caption{Focus of the present research article.}
\label{fig-Paper-Focus}
\end{figure}

This has engendered an appetite for numerical recipes that simplify the PDE into a system of algebraic equations. In CFD, for instance, the process of solving the system of PDEs is depicted in Fig. \ref{fig-Paper-Focus}. It begins by identifying a set of PDEs that describe the observed flow behaviour as the governing equation. Initial conditions and boundary conditions are then specified as dictated by the physics of flow. Since these equations follow the laws of continuum mechanics, the conversion of the governing PDEs to the algebraic system of equations is popularly achieved through grid-based methods, such as the Finite Difference Method (FDM)\cite{ozicsik2017}, Finite Volume Methods (FVM)\cite{moukalled2016} or Finite Element Method (FEM)\cite{huebner2001}. The chosen method determines the manner in which the flow domain is viewed; in FDM, it is distributed collocation points, while in FVM it is interconnected volumes, whereas FEM views the domain as interlinked elements. In transient flow, the strategy employed for time integration often addresses the non-linearity. Explicit method solves each discretized domain one at a time, while the implicit method involves solving the entire system of equations simultaneously, where both the linear and nonlinear terms are treated implicitly. On the other hand, a semi-implicit method combines aspects of both explicit and implicit methods, typically treating the linear terms implicitly and the nonlinear terms explicitly \cite{ascher1997}. Implicit and Semi-implicit methods are generally utilized as it solves the entire domain simultaneously. With these methods, irrespective of the means, the end result is a linear system of equations, which can be expressed as $\textbf{A}x = b$ (Fig. \ref{fig-Paper-Focus}). 

It is crucial to note that nearly all PDE systems can be simplified to the above linear system. Consequently, there have been multi-disciplinary efforts to solve the system, which has led to a plethora of methods, direct (Gaussian Elimination, Thomas Algorithm), iterative (Gauss-Seidel, Gauss Jacobi), and Krylov subspace (Conjugate Gradient, Generalized Minimal Residual) methods \cite{abadir2005}, being developed and utilized over the past several decades. Recent times have also witnessed the rise of code libraries, such as NumPy \cite{ziogas2021}, Intel's MKL\cite{burylov2007}, etc., with significant efforts going into  accelerating the solution to $\textbf{A}x = b$. These methods and implementations have also efficiently handled very large systems. 

However, in certain CFD studies such as a Direct Numerical Solution (DNS) for turbulence \cite{moin1998}, the mesh sizes desired are very small (on the order sub-Kolmogorov scale\cite{schumacher2007}). This leads to massive  $\textbf{A}x = b$ systems that are computationally intractable even with the formidable throughput offered by the modern Exascale supercomputers. 

While the previous generations had the option to wait till computational power increased to meet their need, we do not have that luxury. Modern technology is quickly approaching the limit of Moore's law, physically \cite{powell2008} and economically \cite{rupp2010}. 

In general, the ultimate aim of computational methods is to provide real-time predictions of the complex physical phenomena being modelled. Such a capability elevates such computational tools from merely a design tool kit to a step towards true virtualization. However, such a vision requires immense computational power, which must be a result of a paradigm shift in computing. 

Significant faith is placed in Quantum Computers (QC), which are, theoretically, capable of solving large systems with enormous acceleration \cite{horowitz2019}. Numerous algorithms have been devised for the solution of the Navier-Stokes system of equations, many of which have been discussed in detailed reviews by Gaitan \cite{gaitan2020,gaitan2021}. A few of these approaches are focussed on quantum acceleration of traditional Navier-Stokes solution strategies, such as Finite Difference Methods (FDM), Finite Volume Methods (FVM)\cite{chen2022}, and Finite Element Methods (FEM) \cite{joczik2022}. A larger volume of research has been dedicated to quantum approaches to the Lattice Boltzmann Methods (LBM)\cite{itani2024,succi2024,li2023}. 

Any claim of quantum supremacy can only be established once quantum computers of meaningful capacity have been constructed\cite{arute2019}. Meanwhile, our understanding garnered from existing quantum machines can be gainfully utilized to create hybrid algorithms \cite{endo2021} capable of running on such hardware. The advent of quantum simulators has given an impetus to the quantum algorithm and code development efforts. It is now feasible to test out the efficacy of a computational algorithm on large emulations of quantum hardware \cite{buluta2009,altman2021}, essentially preparing for a time when, inevitably, quantum computers become a reality. Such attempts are historically poetic and reminiscent of Ada Lovelace's efforts to develop code for digital silicon-based computers that didn't exist yet \cite{aiello2016,hollings2018}.

There have been attempts in the quantum field to develop algorithms for the solution of a linear system ($\textbf{A}x = b$) on quantum hardware. One of the original approaches was the Harrow–Hassidim–Lloyd (HHL) algorithm\cite{duan2020} which achieved an exponential acceleration over contemporary classical algorithms. Further improvement on the algorithm's precision was obtained by utilizing Quantum Singular Value Transformations (QSVT)\cite{gilyen2019}, such as the the method proposed by Childs, et. al\cite{childs2017}. However, it is critical to recognize that these speed-ups do not consider the state preparation and state readout requirements, which are themselves, quite challenging. Moreover, these algorithms are meant to be used for fault-tolerant hardware, which is still a few years away. Even those designed for hybrid architectures\cite{lapworth2022} exhibit discouraging performance on quantum simulators in terms of number of qubits (and circuit depth)\cite{yalovetzky2021}. 

Another promising approach, specifically in the Noisy Intermediate-Scale Quantum (NISQ) era, is the Variational Quantum Linear Solver (VQLS) \cite{bravo2023}, which has found remarkable success and has also been utilized to solve heat conduction equation\cite{liu2022} and Poisson equation \cite{liu2021,ali2023}. However, the VQLS algorithm, being a variational algorithm, loses coherence during the computation due to the measurement of the cost function. This loss of coherence limits the potential quantum advantages that can be harnessed. 

This drawback is largely mitigated by the Coherent Variational Quantum Linear Solver (CVQLS), which combines strengths of Variational Quantum Algorithm (VQA) and classical linear algebra. The CVQLS distinguishes itself from the VQLS in three distinct aspects. Firstly, the VQLS utilizes a more general framework for solving optimization problems using quantum computers, while, CVQLS leverages quantum coherence to enhance the accuracy and efficiency of the solution. Secondly, CVQLS encodes the linear system into a quantum state using a coherent process, while VQLS utilizes different embedding techniques. Lastly, the cost function in CVQLS is typically designed for linear systems, while VQLS can use more general cost functions.

Hybrid quantum-classical algorithms are aptly suited to the present NISQ-era and similar methods have been used to solve equations relevant to CFD, such as the Poisson equation \cite{steijl2018,steijl2019}.

This research work utilizes such a hybrid quantum-classical method, in which CVQLS is used as a quantum linear solver algorithm (QLSA), which works well for current quantum computers/simulators. The algorithm, Hybrid Quantum-Classical Finite Methods (HQCFM), can solve $\textbf{A}x = b$ while maintaining high accuracy, consistency and scalability. The HQCFM algorithm give more coherent and precise probability distribution for the solution vector, thereby potentially offering greater quantum speedups.



These claims are put to the test by solving the 2D, transient, incompressible, viscous, non-linear, coupled Burgers equation described in Section \ref{sec2}. Section \ref{sec3} briefly describes the methodology utilised in this research. The accuracy, scalability, and consistency of HQCFM is established in Section \ref{sec4}. Section \ref{sec5} utilizes the knowledge gained in the simulation of the Burgers equation to develop predictive models that provide insight into the cost associated with extending the HQCFM to complex, real-world problems. 
Fig. \ref{fig-Paper-Focus} clarifies this paper's focus and the exact field where HQCFM will make an impact. 

\section{Problem Definition}\label{sec2}
\subsection{Significance of the problem}
One of the most celebrated nonlinear partial differential equations in the field of fluid mechanics is the Burgers equation\cite{bonkile2018}. It is a model equation that is capable of capturing the interaction between advection and diffusion. It arises in gas dynamics, traffic flow, chromatography and flood waves in rivers.

Due to its similarity with the Navier-Stokes equation, it is considered an important model for testing numerical algorithms for the Navier-Stokes equation. The numerical solution of this equation is an indispensable tool for studying incompressible fluid flow problems. 

\subsection{Formulation} \label{formulation}
The Burgers equation is a highly regarded model equation in the field of CFD. It has been shown to model turbulence \cite{srivastava2012} and propagation shockwave in viscous fluids \cite{mukundan2016,mukundan2015}.

The 2D governing equations are, 
\begin{equation}
\frac{\partial u}{\partial t} + u\frac{\partial u}{\partial x}+ v\frac{\partial u}{\partial y} = \frac{1}{Re} \left[\frac{\partial^2 u}{\partial x^2} + \frac{\partial^2 u}{\partial y^2}\right]
\label{eq: u equation}
\end{equation}

\begin{equation}
\frac{\partial v}{\partial t} + u\frac{\partial v}{\partial x}+ v\frac{\partial v}{\partial y} = \frac{1}{Re} \left[\frac{\partial^2 v}{\partial x^2} + \frac{\partial^2 v}{\partial y^2}\right]
\label{eq: v equation}
\end{equation}

For this article, the initial conditions are taken as,
\begin{equation}
\begin{split}
u(x,y,0) &= sin(\pi x) +cos(\pi y) \\
v(x,y,0) &= x+y 
\end{split}
\label{eq: initial condition}
\end{equation}

and the boundary conditions as,
\begin{equation}
\begin{split}
u(0,y,t) = cos(\pi y)  \qquad&\qquad  u(L_x,y,t) = 1 + cos(\pi y) \\
v(0,y,t) = y  \qquad&\qquad  v(L_x,y,t) = 0.5 + y \\
u(x,0,t) = 1 + sin(\pi x)  \qquad&\qquad  u(x,L_y,t) = sin(\pi x) \\
v(x,0,t) = x  \qquad&\qquad  v(x,L_y,t) = 0.5 + x \\
\end{split}
\label{eq: boundary condition}
\end{equation}

\begin{figure}[htbp]
\centering 
\includegraphics[width=\columnwidth]{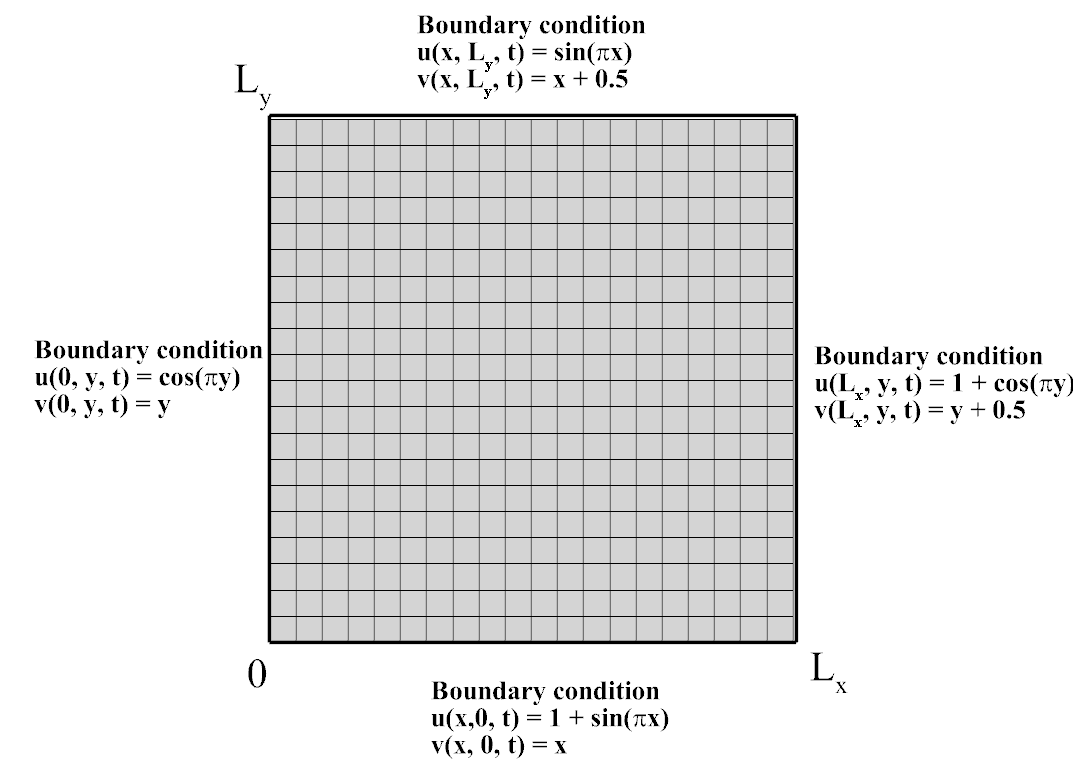}\quad
\caption{Schematic of the computational domain.}
\label{fig-Grid}
\end{figure}

The square domain with $L_x = L_y =0.5 $ is chosen for the study. The numerical computations are performed on a uniform grid whose cell size depends on the chosen grid resolution. The Reynolds number, $Re$, is chosen to be 50. The solution time is chosen to be $0.625$, and time steps are chosen appropriately. The Cartesian discretization and boundary conditions schematic is illustrated in Fig. \ref{fig-Grid}.

The spatial discretization and time integration scheme used by Srivastava and Tamsir \cite{srivastava2012} are chosen here. Studies \cite{you2022,hou2008} demonstrate through stability analysis and numerical experiments that semi-implicit schemes exhibit superior stability properties compared to explicit schemes. While implicit methods solve the entire system simultaneously, semi-implicit methods strike a balance between explicit and implicit treatments of different terms within the equations. Semi-implicit methods are often favored for improved stability, especially when dealing with complex systems involving linear and nonlinear components. Thus, the semi-implicit scheme is used for this study.

The discretization results in the creation of two linear systems corresponding to u and v velocities, which must be solved at each time step. Thus,
\begin{equation}
\begin{split}
[A]{u}^{t+\Delta t} &={u}^{t}\\
[A]{v}^{t+\Delta t} &={v}^{t}\\
\end{split}
\label{eq: linear system}
\end{equation}

The above linear system is solved once for each velocity component. Thus, in a single time step, 2 linear system solutions are obtained.

\begin{figure}[hbt]
\centering
\includegraphics[width=\columnwidth]{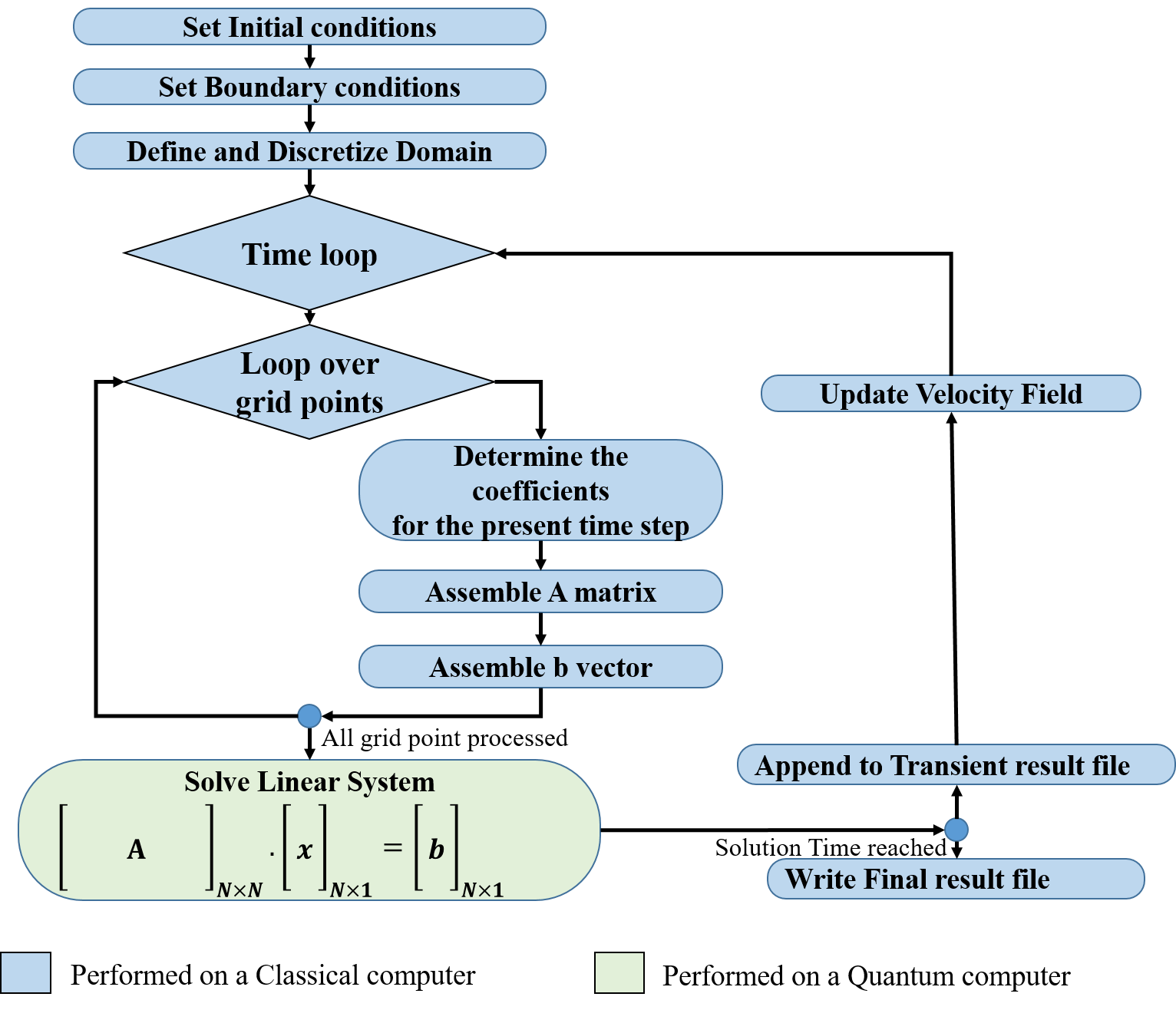}
\caption{Flowchart for the present HQCFM algorithm.} \label{fig-Flowchart_Structure}
\end{figure}

The entire process is outlined in Fig. \ref{fig-Flowchart_Structure}.

\subsection{Setup}

The size of the square matrix $[A]$ is referred to as system size. It is related to the number of grid points in the X and Y direction as, $n_x = n_y =  \sqrt{N}$, where $N$ is the size of the $[A]$ matrix. Thus, a grid of $8 \times 8$ corresponds to a system size of 64. An artifact of HQCFM and other quantum algorithms is the need for the system sizes to be powers of 2. If this is not the case, then padding is utilized to ensure the system meets the criteria. This padding is a numerical fix and doesn't affect or alter the solution in any way. The size of matrices studied in this research is presented in Tab. \ref{SystemsizeTab}.

\begin{table*}[t]
\centering
\caption{\label{SystemsizeTab}Size of the matrix systems used in this study. }
\begin{tabular}{|cccc|}
\hline
\begin{tabular}[c]{@{}c@{}}Actual System Size \\ ($N \times N$)\end{tabular} & \begin{tabular}[c]{@{}c@{}}Grid Resolution \\ ($n_x \times n_y$)\end{tabular} & Padding & \begin{tabular}[c]{@{}c@{}}System Size after Padding\\ ($N \times N$)\end{tabular} \\ \hline
16 x 16                                                               & 4 x 4                                                                         & False   & 16 x 16                                                                            \\
25 x 25                                                               & 5 x 5                                                                         & True    & 32 x 32                                                                            \\
64 x 64                                                               & 8 x 8                                                                         & False   & 64 x 64                                                                            \\
121 x 121                                                             & 11 x 11                                                                       & True    & 128 x 128                                                                          \\
256 x 256                                                             & 16 x 16                                                                       & False   & 256 x 256                                                                          \\
484 x 484                                                             & 22 x 22                                                                       & True    & 512 x 512                                                                          \\
1024 x 1024                                                           & 32 x 32                                                                       & False   & 1024 x 1024                                                                        \\
2025 x 2025                                                           & 45 x 45                                                                       & True    & 2048 x 2048                                                                        \\ \hline
\end{tabular}
\end{table*}

\section{Methodology}\label{sec3}
\subsection{Iterative Classical PDE Solver}\label{secCIS}

The classical solver is used to provide a basis for comparison for the HQCFM. The discretization process and the assembly of the $\textbf{A}x = b$ system remain identical for both methods. The classical and HQCFM perform all the blue boxes in Fig. \ref{fig-Flowchart_Structure}. However, instead of solving the linear system through HQCFM (green box in Fig. \ref{fig-Flowchart_Structure}), the classical solver utilizes an established classical method, namely, the Generalized Miminal RESidual method (GMRES), further details of which can be found in a recent review by Zou\cite{zou2023}.

\subsection{Iterative Quantum PDE Solver}
This work leverages a Quantum Linear System Algorithm (QLSA) well-suited for the NISQ era. We modified an existing QLSA to solve the system of linear equations arising at each time step for semi-implicit discretization of the 2D, transient, incompressible, viscous, non-linear, coupled Burgers equation. As depicted in Fig. \ref{fig-Paper-Focus}, the solution of any partial differential equation (PDE) involves solving a system of the form $\textbf{A}x = b$. We exploit the fact that the $\textbf{A}$ matrix in this system can be decomposed into a linear combination of unitary matrices. This decomposition allows us to embed these matrices efficiently onto a quantum circuit within a Variational Quantum Algorithm (VQA) framework to reach the target state $\ket{\psi}$. The  process of our HQCFM can be summarized as follows, 

\begin{itemize}
   \item  {We decompose our $\textbf{A}$ matrix using spectral theorem as in "Matrix Analysis" by Roger A. Horn and Charles R. Johnson.\cite{horn2013matrix}} 
    \item {Calculating the  Euclidean norm of difference between Quantum $\ket{b}$ and Classical $b$, using this cost for the optimization process.}
    \item {After finding the best $\theta$ (variational) parameters for the circuit, we prepare our $x$ vector. }
    \item {We have utilized IBM Qiskit AerSimulator (GPU-enabled).}
\end{itemize}

\section{Results and Discussion}\label{sec4}

In this section, the accuracy of the HQCFM is established by comparing it with classical solver results. The scalability is then tested by using finer mesh sizes. Finally, comments are made regarding the consistency of the HQCFM by studying the difference between its solution and classical solution at every time step.

\subsection{Accuracy}
\begin{figure*}[!ht]
    \begin{minipage}[l]{0.95\columnwidth}
        \centering
        \includegraphics[width=\textwidth]{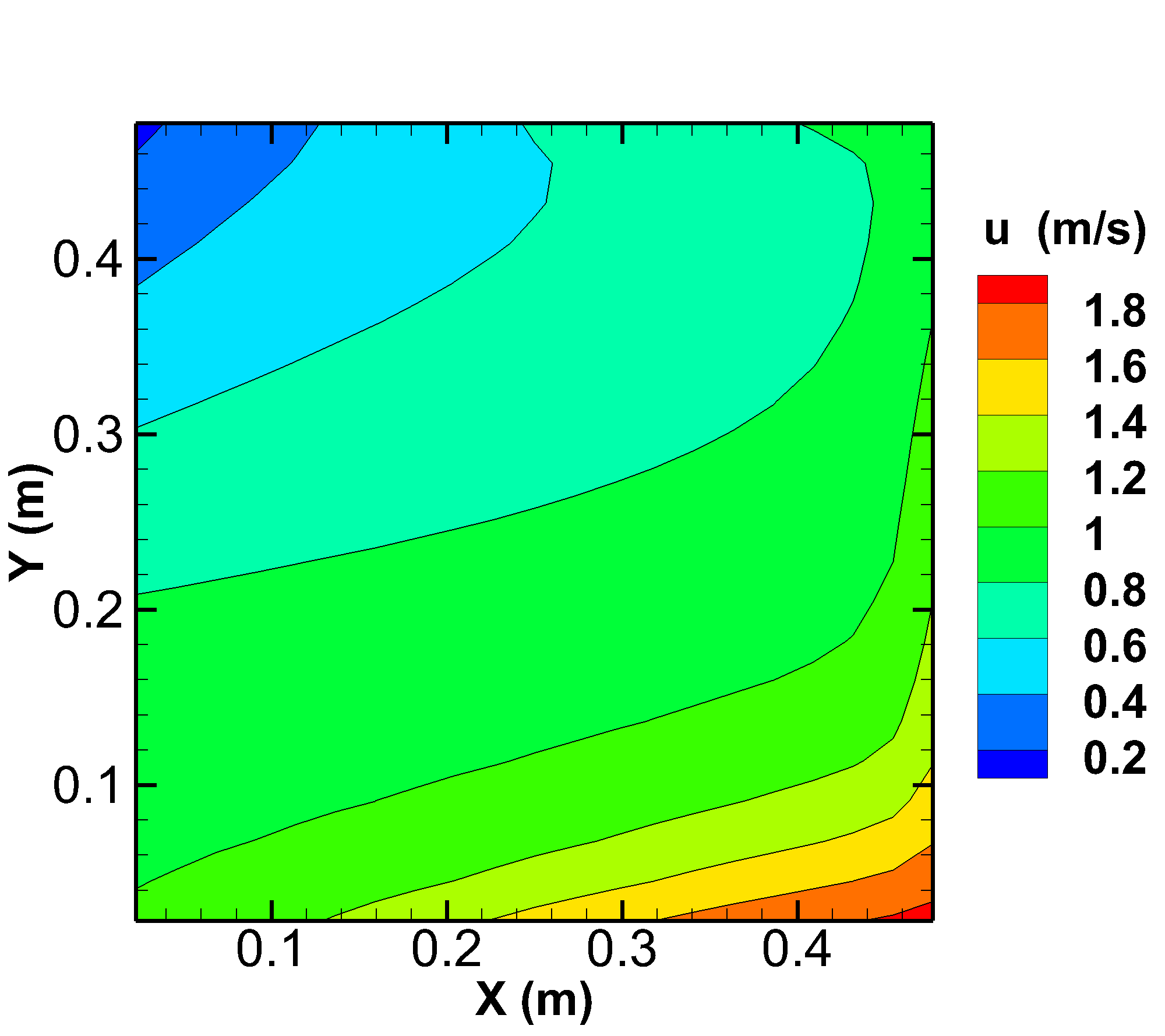}
        \subcaption{$u$-component of velocity}\label{fig:Xvel}
    \end{minipage}
    \hfill{}
    \begin{minipage}[r]{0.95\columnwidth}
        \centering
        \includegraphics[width=\textwidth]{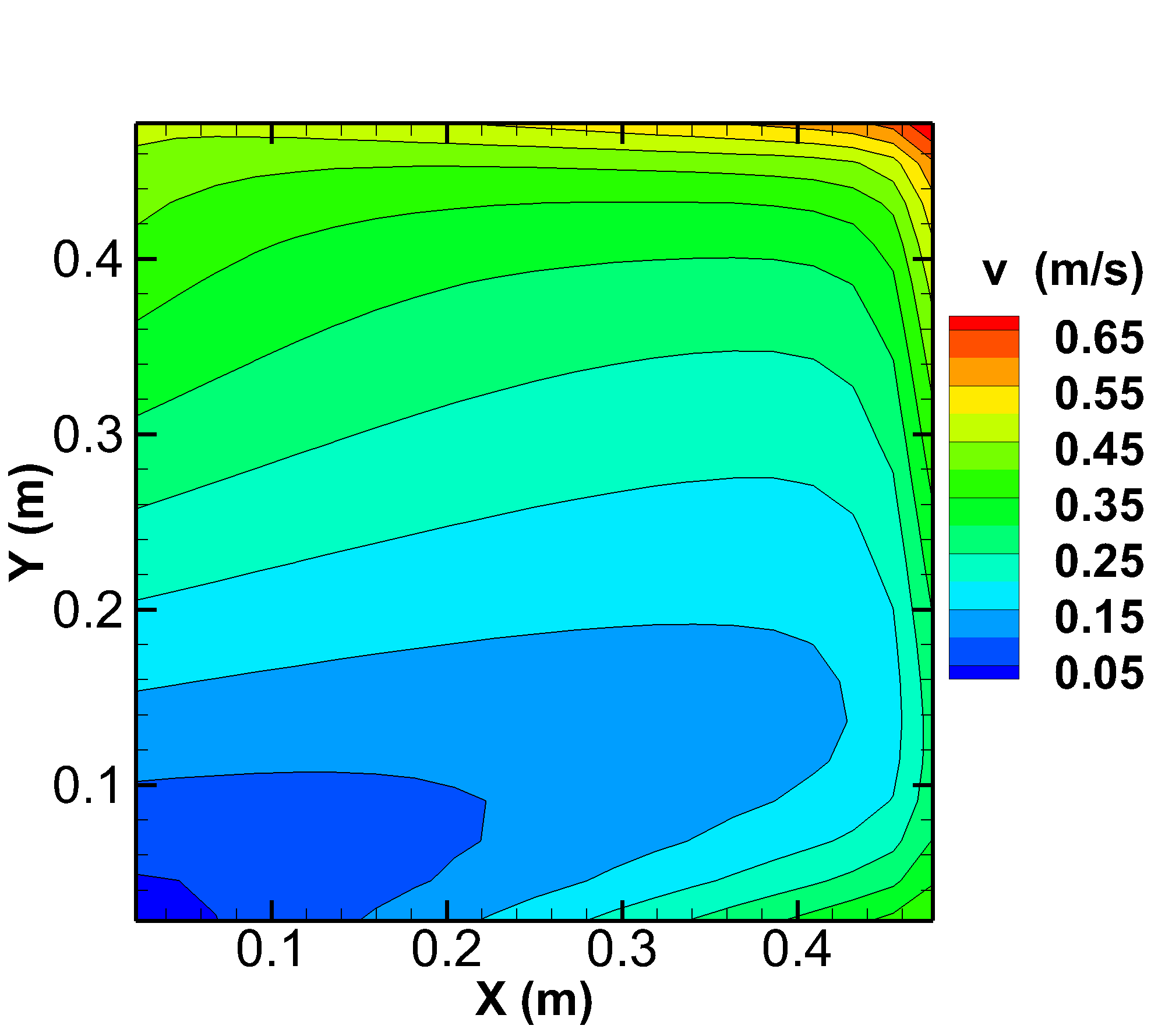}
        \subcaption{$v$-component of velocity}\label{fig:Yvel}
    \end{minipage}
    \hfill{}
    \caption{Contour plots of velocity components and magnitude for a 2D, transient, incompressible, viscous, non-linear, coupled Burgers equation with a 512 $\times$ 512 size system.} \label{fig-UV_CQ512}
\end{figure*}

The velocity contours for the transient solution of the 2D, transient, incompressible, viscous, non-linear, coupled Burgers equation are depicted in Fig. \ref{fig-UV_CQ512}. These results were obtained using the HQCFM introduced in this paper to solve the linear system at every time step. Such a demonstration of using a quantum linear equation solver coupled with a transient problem is unprecedented. Moreover, the results obtained also match the expected flow field for the Burgers’ equation with the initial and boundary conditions specified in the problem statement (refer section \ref{formulation})\cite{srivastava2012}.

\begin{figure*}[!ht]
    \begin{minipage}[l]{0.95\columnwidth}
        \centering
        \includegraphics[width=\textwidth]{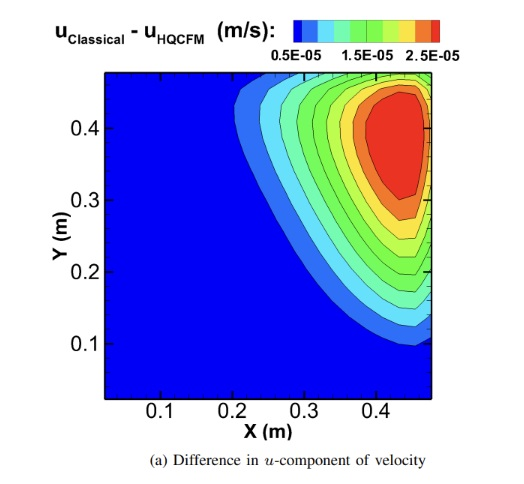}
        \subcaption{Difference in $u$-component of velocity}\label{fig:XvelDiff}
    \end{minipage}
    \hfill{}
    \begin{minipage}[r]{0.95\columnwidth}
        \centering
        \includegraphics[width=\textwidth]{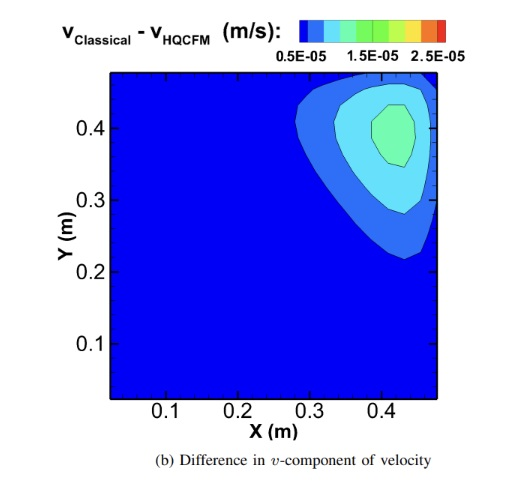}
        \subcaption{Difference in $v$-component of velocity}\label{fig:YvelDiff}
    \end{minipage}
    \hfill{}
    \caption{Contour plots of velocity components and magnitude between classical and quantum solver for a 2D, transient, incompressible, viscous, non-linear, coupled Burgers equation with a 512 $\times$ 512 size system.} \label{fig-UV_CQ512_Difference}
\end{figure*}

 In order to better appreciate the accuracy of HQCFM, we ran identical cases using the classical solver introduced in Section \ref{secCIS}. The resulting contours from the HQCFM and classical solver were subtracted from one another to produce a difference plot as shown in Fig. \ref{fig-UV_CQ512_Difference}. Here, Fig.\ref{fig:XvelDiff} plots the difference in the computation of the $X$ velocity component between classical and the HQCFM, while Fig. \ref{fig:YvelDiff} depicts the difference $Y$ velocity component. 

It can be observed that the maximum disagreement between the two occurs in the top right region of the domain. Although a discrepancy is noted, the difference is very small, on the order of $e^{-5}$. The tables present the physics-based validation of the results, \ref{acc1} and \ref{acc2}. In these tables, the values of the u and v velocity at specific points are compared with those reported in the literature. Minor disagreements are observed, but these can be adequately explained. The discrepancy with respect to Srivastava\cite{srivastava2012} arises because of different grid sizes. The results presented are for a $22 \times 22$ grid, whereas those in the reference are for a $20 \times 20$ grid. On the other hand, the present approach is semi-implicit, whereas that reported in Bahadir\cite{bahadir2003} is fully implicit. The mismatch is consequently, a result of the difference in the treatment of the non-linearity, hence these differences can be attributed to the numerics.

\begin{table*}[t]
\centering
\caption{\label{acc1}Comparison of computed values of u from HQCFM and Classical Solver against literature for $Re=50$ at  $t=0.625$.}
\begin{tabular}{cc|cccc}
X(m) & Y(m) & HQCFM       & Classical   & Srivastava\cite{srivastava2012} & Bahadir\cite{bahadir2003} \\ \hline
0.1  & 0.1  & 0.962570961 & 0.962571185 & 0.97146           & 0.96688        \\
0.3  & 0.1  & 1.096567222 & 1.096580234 & 1.1528            & 1.14827        \\
0.2  & 0.2  & 0.861248176 & 0.861256984 & 0.86307           & 0.85911        \\
0.4  & 0.2  & 0.939141661 & 0.939266899 & 0.97981           & 0.97637        \\
0.1  & 0.3  & 0.664038717 & 0.664042106 & 0.66316           & 0.66019        \\
0.3  & 0.3  & 0.764007634 & 0.76410028  & 0.7723            & 0.76932        \\
0.2  & 0.4  & 0.581100144 & 0.581146485 & 0.5818            & 0.57966        \\
0.4  & 0.4  & 0.748058121 & 0.748329615 & 0.75855           & 0.75678        \\ \hline
\end{tabular}
\end{table*}

\begin{table*}[t]
\centering
\caption{\label{acc2}Comparison of computed values of v from HQCFM and Classical Solver against literature for $Re=50$ at  $t=0.625$.}
\begin{tabular}{cc|cccc}
X(m) & Y(m) & HQCFM       & Classical   & Srivastava\cite{srivastava2012} & Bahadir\cite{bahadir2003} \\ \hline
0.1  & 0.1  & 0.093884394 & 0.093884452 & 0.09869           & 0.09824        \\
0.3  & 0.1  & 0.118062887 & 0.118066365 & 0.14158           & 0.14112        \\
0.2  & 0.2  & 0.16395324  & 0.163956008 & 0.16754           & 0.16681        \\
0.4  & 0.2  & 0.157233779 & 0.15727113  & 0.17109           & 0.17065        \\
0.1  & 0.3  & 0.264646384 & 0.264647809 & 0.26378           & 0.26261        \\
0.3  & 0.3  & 0.224208984 & 0.224244041 & 0.22654           & 0.22576        \\
0.2  & 0.4  & 0.314125289 & 0.314146998 & 0.32851           & 0.32745        \\
0.4  & 0.4  & 0.305482011 & 0.305587781 & 0.32499           & 0.32441        \\ \hline
\end{tabular}
\end{table*}

\subsection{Scalability}

\begin{figure*}[!ht]
    \begin{minipage}[l]{\columnwidth}
        \centering
        \includegraphics[width=0.95\textwidth]{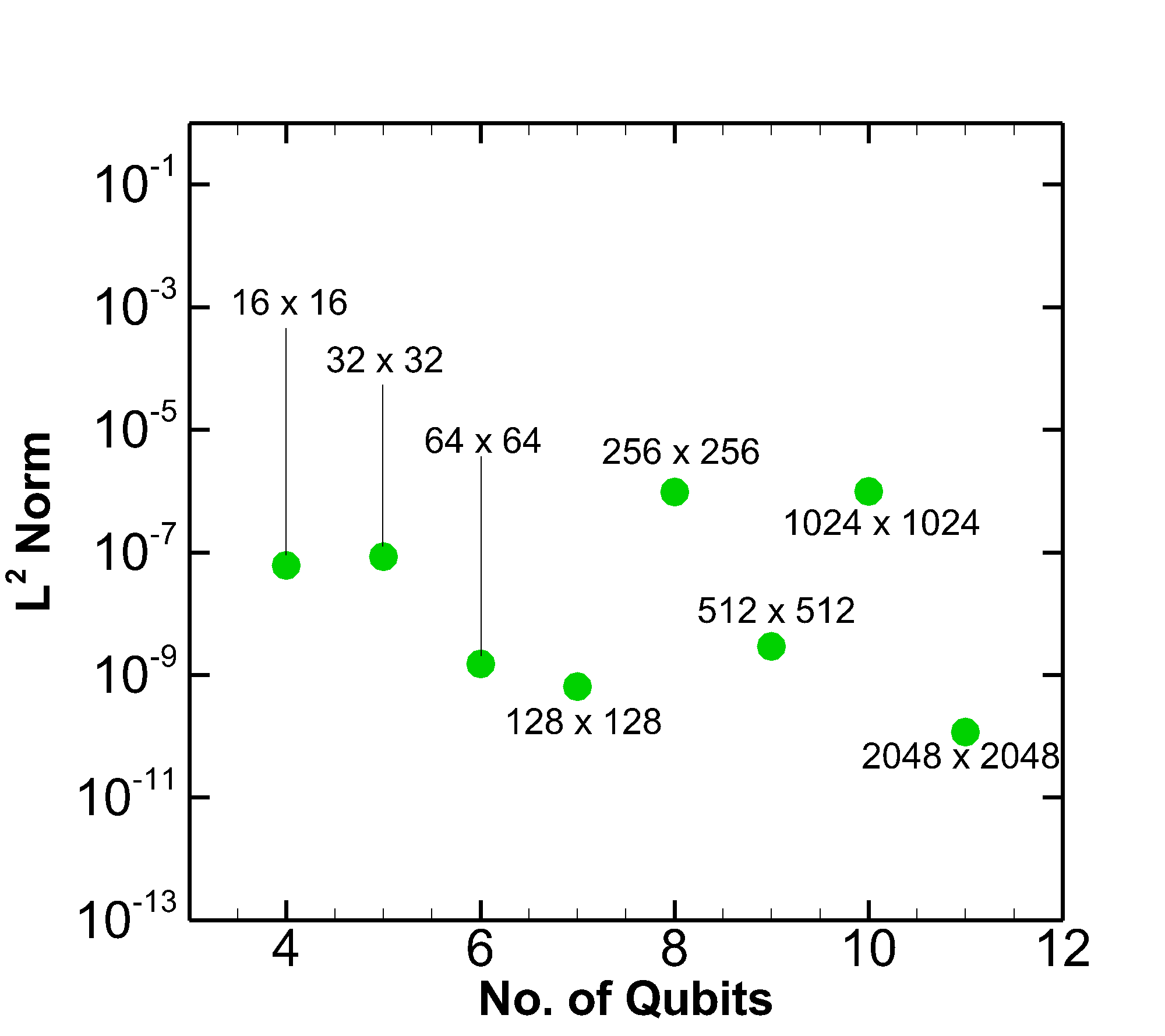}
        \subcaption{Variation of accuracy for various systems sizes w.r.t. number of qubits}\label{fig:AAA}
    \end{minipage}
    \hfill{}
    \begin{minipage}[r]{\columnwidth}
        \centering
        \includegraphics[width=0.95\textwidth]{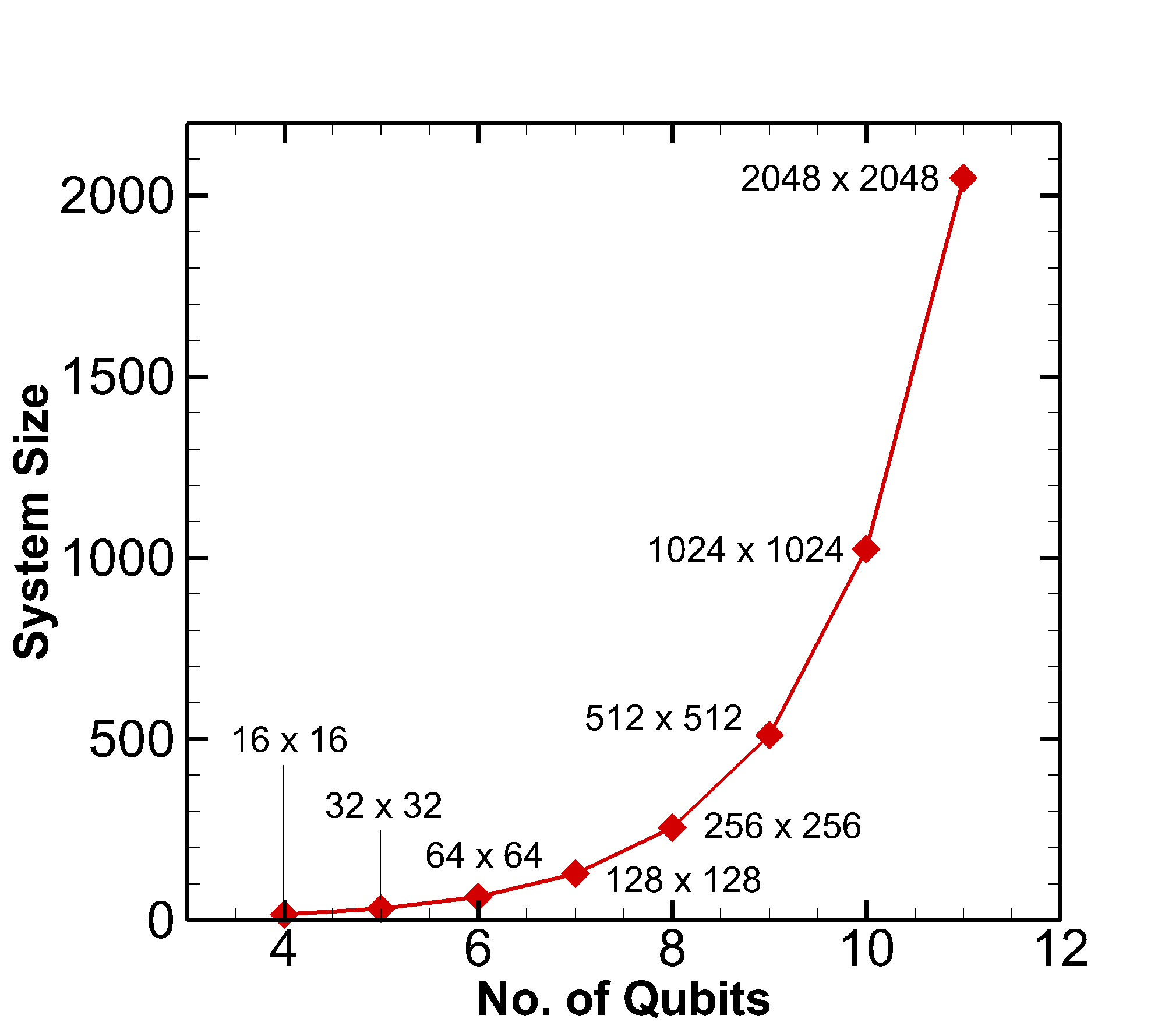}
        \subcaption{Scalability of system size with number of qubits}\label{fig:BBB}
    \end{minipage}
    \caption{Demonstration of scalability and accuracy of HQCFM for a 2D, transient, incompressible, viscous, non-linear, coupled Burgers equation.}\label{fig-scalability&accuracy}
\end{figure*}

Although the accuracy of the quantum linear system solver with regards to the classical solver has been established for a $512 \times 512$ sized system, this is not the top limit. In fact, a number of other simulations with variations in the system size have also been performed to gauge the scalability properties of the HQCFM. The accuracy of these simulations is presented in terms of the L2 norm between the HQCFM and classical results in Fig. \ref{fig:AAA}. It is observed that the L2 norm is in the range of $1e^{-5}$ to $1e^{-11}$, nearing machine zero, even for systems as large as 2048. This provides enormous confidence in the ability of the developed quantum solver to scale up to larger problems.

Accuracy alone is not the only factor affecting scalability. Quantum algorithms require a number of qubits to be held in a superposition state, which is a physically arduous process. The number of qubits increases with the system size. In the present research, the number of qubits used by the algorithm for each of the systems is measured and plotted in Fig. \ref{fig:BBB}. A more robust breakdown of the number of qubits and quantum gates is provided in Tab. \ref{tab:1}. 

\subsection{Consistency}

\begin{figure}[hbt]
\centering
\includegraphics[width=0.95\columnwidth]{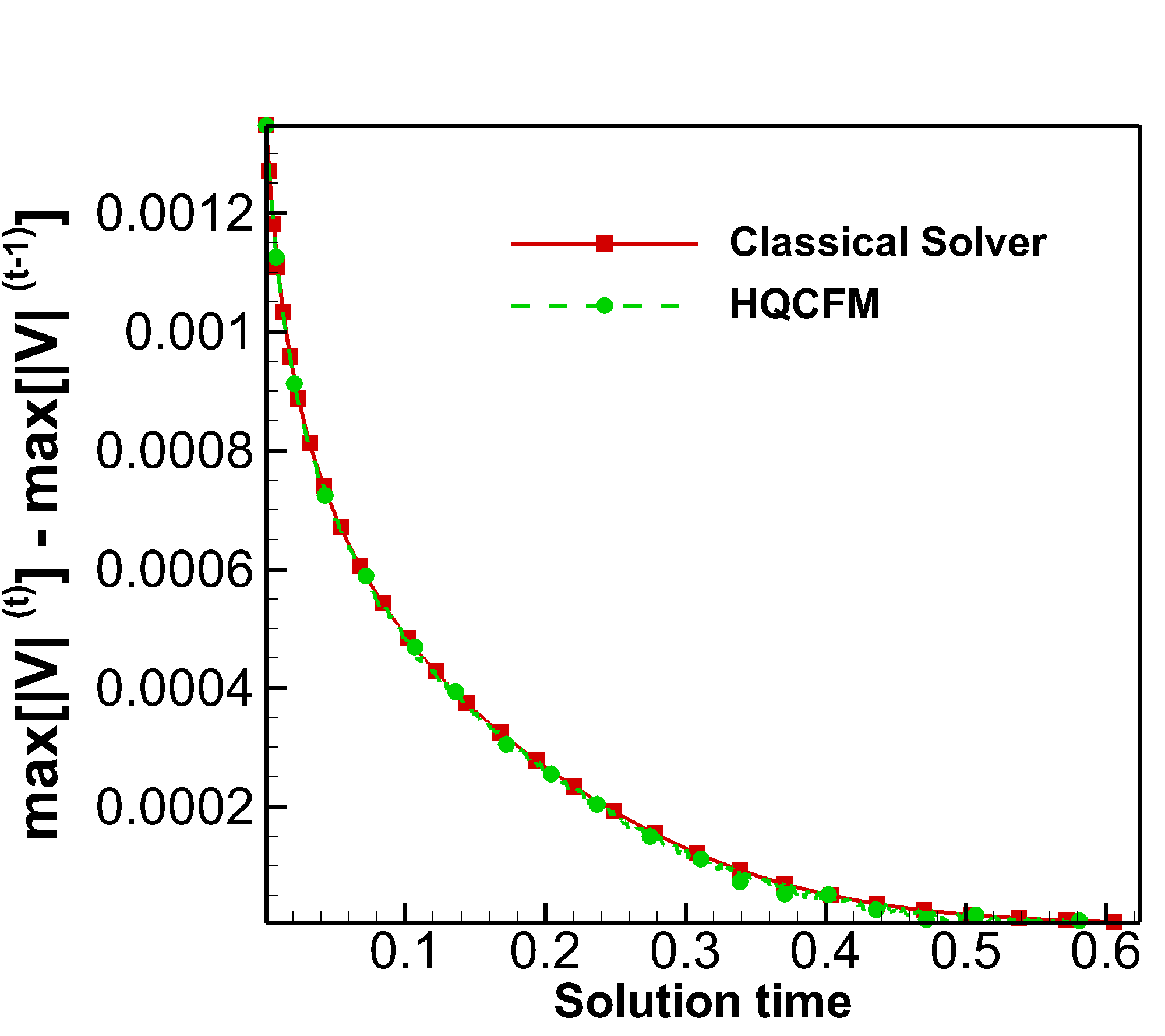}
\caption{Variation of maximum magnitude of velocity in consecutive time steps with solution time for quantum and classical solvers.} \label{fig-residual}
\end{figure}

It is worth reiterating that this article demonstrates the developed capability of HQCFM, achieving not only high accuracy and scalability but also exhibiting remarkable consistency in these features. Notably, the quantum solver is integrated into the time loop, requiring evaluation at every time step for the Burgers' equation. This equation's non-linear nature necessitates re-evaluation of the coefficient matrix, $\textbf{A}$, based on the previous step's results at each iteration.
This emphasizes the critical need for the HQCFM to maintain consistent accuracy throughout the simulation, as any errors can propagate and significantly alter the trajectory toward the steady-state solution. However, as Fig. \ref{fig-residual} illustrates,  the solutions obtained using the HQCFM and the classical solver exhibit near-identical convergence behavior towards the steady state. This consistency highlights the effectiveness of the HQCFM in maintaining accuracy even under the iterative nature of the time-stepping algorithm.

\section{Resource Estimation} \label{sec5}

The resources utilized by the HQCFM are listed in Tab. \ref{tab:1} for the simulations performed in this research. These are crucial for trying to predict resources required for larger simulation systems.

\begin{table*}[t]
\caption{Resources utilized by the HQCFM quantum circuit for different system sizes. Value in the brackets is the no. of Cx gates out of the total multi-qubit gates.}
\centering
\begin{tabular}{|c|cccc|}
\hline
Size of   the System & No. of   Qubits & No. of Single   Qubit Gates & No. of Multi-Qubit Gates (No. of Cx gates) & Total Gates \\ \hline
16                   & 4               & 76                          & 42  (Cx = 36)                         & 118         \\
32                   & 5               & 95                          & 66  (Cx = 60)                         & 161         \\
64                   & 6               & 114                         & 96  (Cx = 90)                         & 210         \\
128                  & 7               & 133                         & 132 (Cx = 126)                        & 265         \\
256                  & 8               & 152                         & 174 (Cx = 168)                        & 326         \\
512                  & 9               & 171                         & 222 (Cx = 216)                        & 393         \\
1024                 & 10              & 190                         & 276 (Cx = 270)                        & 466         \\
2048                 & 11              & 209                         & 336 (Cx = 330)                        & 545         \\ \hline
\end{tabular}
  \label{tab:1}
\end{table*}

Consider a complex problem like internal flow in turbomachinery, which is characterized by non-linearity and turbulence. Such a flow can only be resolved by high mesh sizes and significant computational power. For a mesh of size $2^{30}$, or approximately 1.69 billion cells, a traditional high-performance computer (HPC) system would require about 19.2 million cores \cite{fu2023}.

When the HQCFM presented in this research is run on an ideal quantum computer, this same problem could be tackled using only 30 qubits. However, it is also paramount that we have an estimate of the Quantum resources the HQCFM would utilize to run such a simulation. Such a resource estimation exercise is performed here using an arithmetic progression model based on the data gathered in Tab. \ref{tab:1}.

An arithmetic progression model can estimate the number of single qubit gates, as depicted in Eq. \ref{eq: single qubit estimation}.

\begin{equation}
a_n = a_1 + (n-1)\,d
\label{eq: single qubit estimation}
\end{equation}

Here, $a_n$ is the number of single qubit gates required for simulating the target system size,  $a_1$ is the starting point, which is taken to be the $16 \times 16$  system. Thus, by consulting the table $a_1 = 76$. Additionally, $n$ is the number of qubits, and $d$ is the difference between two consecutive systems, which is obtained by taking the difference between consecutive systems' single qubit gate requirement. From consulting the tab. \ref{tab:1}, $d = 19$.

Thus, for a $n=30$, the number of single qubits gates utilized by HQCFM would be $ 76 + (29) \times 19 = 627 $

The number of multi-quit gates can also be estimated by an arithmetic progression model, as depicted in Eq. \ref{eq: multi qubit estimation-1}. 

\begin{equation}
\begin{split}
d_n &= d_1 +  (n-1)\, d \\
a_n &= a_1 + \sum_{k=1}^{n-1}\, d_k\\
\end{split}
\label{eq: multi qubit estimation-1}
\end{equation}
After rewriting

\begin{equation}
a_n = a_1 + \frac{(n-1) (d_1+d_n)}{2}
\label{eq: multi qubit estimation-2}
\end{equation}

From consulting Tab. \ref{tab:1}, the value of $a_1 = 42$, $d_1 = 24$, and the common difference of differences, $d = 6$. Here, $d_n$ is computed first using Eq. \ref{eq: multi qubit estimation-1}, as $d_n = 24 + (29) \times 6 = 198$. Thus, for a $n=30$, the number of multi-qubits gates utilized by HQCFM would be obtained from Eq. \ref{eq: multi qubit estimation-2} as, $ a_n = 42 + (29) * (24 + 198)/2 = 3261 $.

Thus, it would take 627 single-qubit gates and 3261 multi-qubit gates, totalling 3882 gates, for the HQCFM to perform the internal, turbulent flow inside a turbomachinery device using a grid with 1.69 billion cells. Thus, the HQCFM offers a reduction in hardware requirements compared to traditional HPC while also enabling highly complex simulations.  

\section{Conclusion}

The rising demand for true virtualization, either through digital twins or real-time solutions, combined with the surging size and complexity of the associated Partial Differential Equations (PDEs) has challenged and outpaced the existing High Performance Computers. With the onset of the Moore limit, the hope that silicon-based computers can address these need is fading. Next-generation technologies are needed to satisfy current demand and sustain the course of scientific enquiry into the future. Quantum computing, is one of the few options that shows tremendous promise. 

In this article, the Hybrid Quantum-Classical Finite Method (HQCFM) is developed based on an existing Quantum Linear Solver Algorithm (QLSA). The focus of the HQCFM was to solve the $\textbf{A}x=b$ linear system, which is utilized in the present research to solve 2D, transient, incompressible, viscous, non-linear, coupled Burgers equation. Such a demonstration of using a quantum linear equation solver coupled with a transient problem is unprecedented. The performance of HQCFM is measured using 3 metrics, viz., Accuracy, Scalability, and Consistency, 

Present results indicate a very high accuracy when compared to classical solvers. The results are also verified with those reported in the literature. Various system sizes are used to judge the scalability of HQCFM. The results indicate a good scale-up behavior. Quantitative information regarding the scale-up is also provided in terms of qubits and quantum gates utilized. The HQCFM also distinguishes itself by running inside a time loop in a transient problem, without propagating any error to the next time step. Obtaining such a high accuracy consistently in a system of linked linear systems is a major breakthrough in the field of quantum computing.

Apart from these scientific findings, this paper also explores the economical aspect of quantum computing. A robust resource estimation, based on information gathered in this research, is used to predict the resources needed by HQCFM to perform complex real-world simulations. These predictions indicate that a 30-qubit quantum computer will be able to outperform an Exascale computer.

\section*{Acknowledgment}

The authors would like to thank Mr. Abhishek Singh (HPC Engineer at BosonQ Psi) and Mr. Aman Mittal (HPC Engineer at BosonQ Psi) for their support with computational resources; Mr. Akshith Chowdary (Quantum Software Developer at BosonQ Psi) and Mr Aakif Akthar (Quantum Algorithm Developer at BosonQ Psi)  for their keen insights and oversight during the writing process; Dr. Kandula Eswara Sai (Senior Optimization Scientist at BosonQ Psi) for guidance and support.

\bibliographystyle{unsrt}
\bibliography{References}
\end{document}